\documentstyle[twocolumn,aps,epsfig]{revtex}
\begin{document}
\draft
\title{
Exact Numerical Calculation of the Density of States
of the Fluctuating Gap Model}

\author{Lorenz Bartosch and Peter Kopietz} 
\address{
Institut f\"{u}r Theoretische Physik, Universit\"{a}t G\"{o}ttingen,
Bunsenstrasse 9, D-37073 G\"{o}ttingen, Germany}
\date{August 4, 1999}
\maketitle
\begin{abstract}
We develop a powerful numerical algorithm for
calculating the density of states $\rho ( \omega )$ 
of the fluctuating gap model,
which describes the low-energy physics of
disordered Peierls and spin-Peierls chains.
We obtain 
$\rho ( \omega )$
with unprecedented accuracy
from the solution of a simple {\it{initial value problem}}
for a single Riccati equation. 
Generating  Gaussian disorder with large correlation length $\xi$ by
means of 
a simple Markov process, we present a quantitative study
of the behavior of $\rho ( \omega )$ in the pseudogap regime.
In particular, we show that in the commensurate case
and in the absence of forward scattering
the pseudogap is overshadowed by a Dyson singularity
below a certain energy scale $\omega^{\ast}$, which we explicitly
calculate as a function of $\xi$.
\end{abstract}
\narrowtext
%
%
%

The fluctuating gap model (FGM) describes the low-energy
physics of one-dimensional
fermions subject to static disorder
potentials. 
The first quantized Hamiltonian 
of the FGM can be written as\cite{Bunder99}
 \begin{equation}
 \hat{H}  =   
 -  i v_F  \partial_x \sigma_3 + V ( x ) \sigma_0
 +
 \Delta ( x ) \sigma_{+} + \Delta^{\ast} ( x ) \sigma_{-} 
 \label{eq:Hamiltonian}
 \; ,
 \end{equation}
where $V  ( x )$ and $\Delta ( x )$  are random potentials
describing forward and backward scattering,
$v_F$ is the Fermi velocity (henceforth we set $v_F = 1$), 
$\sigma_{i}$ are the usual Pauli matrices, $\sigma_0$ is the
$2 \times 2$ unit matrix, and
$\sigma_{\pm} = \frac{1}{2}(\sigma_{1} \pm i \sigma_{2})$.
Eq.(\ref{eq:Hamiltonian}) emerges as the effective low-energy
Hamiltonian in different physical contexts.
For example, fluctuation effects close to the Peierls
transition in quasi-one-dimensional charge-density wave 
systems can be described by this Hamiltonian.
In this case $\Delta ( x )$ describes the time-independent
part of the fluctuating Peierls order parameter,
the probability distribution of which
can be obtained from  a Ginzburg-Landau expansion
of the free energy\cite{Lee73}.
For commensurate chains 
$\Delta ( x )$ can be chosen to be real,
whereas it is complex in the incommensurate 
case\cite{Bunder99,Lee73,Tchernyshyov98}.
Truncating the Ginzburg-Landau expansion at the second order,
$\Delta ( x )$ is approximated by a Gaussian random process,
with finite average
$\langle \Delta ( x ) \rangle = \Delta_{\rm av}$ below the Peierls
transition and 
$\Delta_{\rm av} = 0$ in the disordered phase.
For commensurate chains the correlator of 
$\tilde{\Delta} ( x ) = \Delta ( x ) - \Delta_{\rm av}$ is
$\langle \tilde{\Delta} ( x ) \tilde{\Delta} ( x^{\prime} )  \rangle
= \Delta_s^2 e^{- | x - x^{\prime} | / \xi}$,
whereas in the incommensurate case
$\langle \tilde{\Delta} ( x ) \tilde{\Delta}^{\ast} ( x^{\prime} )  \rangle
= \Delta_s^2 e^{- | x - x^{\prime} | / \xi}$ and
$\langle \tilde{\Delta} ( x ) \tilde{\Delta} ( x^{\prime} )  \rangle = 0$.
Here $\xi$ is the order parameter correlation length, which 
diverges at the Peierls transition.
The Hamiltonian (\ref{eq:Hamiltonian}) describes also
the low-energy physics of disordered spin chains\cite{Bunder99,Fabrizio97},
which can be mapped onto disordered fermions by means
of the usual Jordan-Wigner 
transformation.
In many cases the filling of the effective fermionic system
is then commensurate with the lattice, so that 
$\Delta ( x )$ is real.

The fundamental quantity which determines
the thermodynamics of the model (\ref{eq:Hamiltonian})
is the density of states (DOS) $\rho ( \omega )$.
In general, one has to rely on approximations
to calculate $\rho ( \omega )$  or its disorder average
$\langle \rho ( \omega ) \rangle$, but
in special limits exact results 
are available. Besides the trivial case where
$V ( x)$ and $\Delta ( x )$ are constant, 
the exact $\langle \rho ( \omega ) \rangle$ can be
obtained by various  
methods\cite{Ovchinnikov77,Lifshits88,Hayn96}
in the white noise limit
$\xi \rightarrow 0$, $\Delta_s \rightarrow \infty$, with 
$\Delta_s^2 \xi \rightarrow  {\rm const}$. 
For real $\Delta ( x )$ and  $V ( x ) =0$ 
the average DOS is known to exhibit,
for sufficiently small $\Delta_{\rm av}$,
a Dyson singularity\cite{Dyson53} at $\omega = 0$.
In Ref.\cite{Bartosch99a} we have shown that this singularity
survives for arbitrary $\xi < \infty$. 
A recursive algorithm due to 
Sadovskii\cite{Sadovskii79} does not
reproduce the Dyson singularity, 
so that this algorithm cannot be exact.
In fact, a subtle flaw in this algorithm has 
recently been found by Tchernyshyov\cite{Tchernyshyov98}.
Because Sadovskii's algorithm (or generalizations of it)
has been used in different contexts, e.g.
to explain the pseudogap phenomenon in the normal state
of the cuprate superconductors\cite{Schmalian98,McKenzie96}, 
it is also important to investigate
its validity
for complex $\Delta (x )$
with quasi-long-range correlations.

In this work we develop an accurate algorithm which allows us to
investigate the regime 
$\Delta_s \xi 
{ \raisebox{-0.5ex}{$\; \stackrel{>}{\sim} \;$}}  1$
where no exact solution is available. 
We find that in the commensurate case the pseudogap is overshadowed by
a Dyson singularity below a cross-over energy $\omega^{\ast}$, which
we determine as a function of the correlation length $\xi$. We also
consider the incommensurate case for which Sadovskii's solution turns
out to be qualitatively correct but leads to a wrong
$\xi$-dependence of the depth of the pseudogap.

{\it{Riccati equation.}}--
In the following we
use the special symmetries of the 
continuum model (\ref{eq:Hamiltonian}) to
show that the DOS can be obtained {\it{without ever calculating
the eigenvalues of $\hat{H}$}}\cite{Millis99}.
Instead, we obtain the DOS
from the solution of a simple {\it{initial value problem}}
for a Riccati equation. This will enable us to calculate
$\rho ( \omega )$ with unprecedented numerical accuracy.
For a given realization of the disorder the local DOS 
of the Hamiltonian (\ref{eq:Hamiltonian}) 
can be defined by\cite{Bartosch99a,footnote1}
\label{eq:footnote1}
 \begin{equation}
 \rho ( x , \omega ) = - {\pi}^{-1}
 {\rm Im} {\rm Tr} [ \sigma_3
 {\cal{G}}^{R} ( x , x , \omega ) ]
 \; ,
 \label{eq:localdos}
 \end{equation}
where the retarded $2 \times 2$ matrix Green function 
${\cal{G}}^{R} ( x , x' , \omega)$
satisfies
 \begin{eqnarray}
 [ i \partial_x - M(x,\omega + i0^{+}) ]
 {\cal{G}}^{R} ( x , x^{\prime} , \omega)
 = \delta ( x - x^{\prime}) \sigma_0
 \label{eq:matrix}
 \; , \\
  \label{M}
  M (x,\omega) = [V(x) -\omega + \Delta ( x ) \sigma_{+} + \Delta^{\ast}
  (x) \sigma_{-} ] \sigma_{3}
  \; .
\end{eqnarray}
We now make the non-Abelian Schwinger-ansatz\cite{Bartosch99a,Schwinger62}
 \begin{equation}
 {\cal{G}}^{R} (x,x',\omega  ) = U (x,\omega)
 {\cal{G}}_{0}^{R} ( x - x'  )  U^{-1} (x',\omega)
 \; ,
 \label{eq:ansatz}
 \end{equation}
where $U (x,\omega)$ is an invertible $2 \times 2 $ matrix and
${\cal{G}}_{0}^{R} (x)$ is the Green function to
the operator $i\partial_{x} +i0^{+}\sigma_{3}$, i.e.
\begin{equation}
  \label{G_0^R}
  {\cal{G}}^{R}_{0} (x) = -i 
  \left( \begin{array}{cc}
   \theta(x) & 0 \\ 
   0 & -\theta(-x) 
 \end{array} \right) \; . 
\end{equation}
In the following the
$\omega$-dependence is suppressed.
The ansatz (\ref{eq:ansatz}) indeed solves Eq.(\ref{eq:matrix})
if $U(x)$ satisfies
 \begin{equation}
 \left[ i \partial_x - M(x) \right] U(x) = 0
 \label{eq:Udif}
 \end{equation}
with the boundary conditions
\begin{equation}
  \label{BC}
  U_{12}(-\infty) = U_{21}(\infty) = 0 \; .
\end{equation}
Two different solutions of Eq.(\ref{eq:Udif}) are given by
\begin{eqnarray}
  \label{U+}
  U_{+}(x) & = &
 \text{Texp}[-i\int_{-\infty}^{x} M(y) dy] 
 \; ,
  \label{U-} \\
  U_{-}(x) & = &
\text{T}^{-1}\text{exp}
  [i\int_{x}^{\infty} M(y) dy]  \; ,
\end{eqnarray}
where Texp is the path-ordered and $\text{T}^{-1}\text{exp}$ is the
anti-path-ordered exponential function. 
Because 
$M^{\dagger} = \sigma_3 M \sigma_3$ and
${\rm Tr} M = 0$,
the matrices $U_{\alpha}$ satisfy
$U_{\alpha}^{\dagger}  = \sigma_3 U_{\alpha}^{-1} \sigma_3$ and
${\rm det} U_{\alpha} =1$, which means
that they belong to the
non-compact group $SU(1,1)$.  
Thus, the elements of the $U_{\alpha}$ satisfy
$U_{\alpha22}=U_{\alpha11}^{\ast}$,
$U_{\alpha12}=U_{\alpha21}^{\ast}$, and
$| U_{\alpha 11}|^2 - 
| U_{\alpha 21}|^2  =1 $.
While each 
$U_{\alpha}(x)$ only obeys one of the two conditions (\ref{BC}), 
the combination
\begin{equation}
  \label{eq:U}
  U(x) \equiv \frac{1}{\sqrt{u}}\left( \begin{array}{cc}
   U_{-11}(x) & U_{+12}(x) \\ 
   U_{-21}(x) & U_{+22}(x)
 \end{array} \right) \; ,
\end{equation}
satisfies both boundary conditions. 
Here $u = U_{-11}(-\infty) = U_{+22}(\infty)$, so that
${\rm det} U(x) = 1$. 
Denoting the first
column of the matrix $U_{\alpha}$ by ${\bf u}_{\alpha}$ and the second
column by ${\bf v}_{\alpha}$ (so that ${\bf v}_{\alpha} = \sigma_{1}
{\bf u}_{\alpha}^{\ast}$), we obtain
from Eqs.(\ref{eq:ansatz}) and (\ref{eq:U})
\begin{eqnarray}
  \label{Greenfunction}
   {\cal{G}}^{R} ( x , x^{\prime} , \omega) = -i\left\{ \theta(x-x')
       \frac{{\bf{u}}_{-}(x) {\bf{u}}_{+}^{\dagger}(x')}{u} \right. \nonumber \\
     + \left. \theta(x'-x)
       \frac{{\bf{v}}_{+}(x) {\bf{v}}_{-}^{\dagger}(x')}{u} \right\} \sigma_{3}\; .
\end{eqnarray}
Here ${\bf{u}}_{+}^{\dagger}$ constitutes adjungation of
${\bf{u}}_{+}$, so that ${\bf{u}_{-}}
{\bf{u}}_{+}^{\dagger}$ is a $2 \times 2$-matrix.
Equivalent but more complicated forms of Eq.(\ref{Greenfunction}) 
were first
derived by Abrikosov and Ryzhkin\cite{Abrikosov76}.
Combining Eqs.(\ref{eq:localdos}) and (\ref{Greenfunction}), we get
\begin{equation}
  \rho(x,\omega) = \frac{1}{\pi} \text{Re} \frac{U_{-11}
  U_{+11}^{\ast} + {U_{-21}} U_{+21}^{\ast}}{U_{-11}
  U_{+11}^{\ast} - {U_{-21}} U_{+21}^{\ast}} \;.
\end{equation}
Since this expression only depends on the ratios $\Phi_{\alpha}(x)
\equiv -i
U_{\alpha21}^{\ast}(x)/U_{\alpha11}^{\ast}(x)$,
we may also write 
\begin{equation}
  \label{localDOSRic}
  \rho(x,\omega) = \frac{1}{\pi} \text{Re}
  \frac{1+\Phi_{+}(x)\Phi_{-}^{\ast}(x)}{1-\Phi_{+}(x)\Phi_{-}^{\ast}(x)} \; .
\end{equation}
Using  Eq.(\ref{eq:Udif})  we find that
the $\Phi_{\alpha}(x)$ are both solutions of the same Riccati equation,
\begin{equation}
  \label{Riccati}
  \partial_{x} \Phi_{\alpha}(x) =
  2i\tilde \omega(x) \Phi_{\alpha}(x) + \Delta(x) - \Delta^{\ast}(x)
  \Phi_{\alpha}^{2}(x)  \; ,
\end{equation}
where we have introduced $\tilde \omega(x) = \omega - V(x)$.
Similar Riccati equations have
recently been obtained by Schopohl\cite{Schopohl98} from the
Eilenberger equations of superconductivity.
To specify the initial conditions, let us assume that
outside the interval $[0, L]$  the potentials
$V(x)$ and $\Delta(x)$ are real constants, 
$V_{\infty}$ and
$\Delta_{\infty}$. From the definition of
$\Phi_{\alpha}$ we find that 
Eq.(\ref{Riccati}) should then be integrated with the initial conditions
\begin{equation}
  \Phi_{+} (0 ) = \Phi_{-} ( L ) =
  \sqrt{1-\frac{(\omega-V_{\infty})^{2}}{\Delta_{\infty}^{2}}} +
  i \frac{\omega-V_{\infty}}{\Delta_{\infty}} \; ,
  \label{initial}
\end{equation}
where the square root has to be taken such that 
for $\Delta_{\infty} \rightarrow 0$ the
right-hand side of Eq.(\ref{initial}) vanishes.
Note that the initial values are
simply given by the stable stationary solution of the Riccati equation
(\ref{Riccati}) with $V(x) = V_{\infty}$ and $\Delta(x) = \Delta_{\infty}$.

{\it{The case of a discrete spectrum.}}--
For $(\omega-V_{\infty})^{2} < \Delta_{\infty}^{2}$ 
the spectrum turns out to be discrete\cite{footnoteLetter}:
Introducing $\varphi_{\alpha}(x)$ via $\Phi_{\alpha}(x) \equiv
e^{i\varphi_{\alpha}(x)}$ 
the phases satisfy
\begin{equation}
  \label{eq:phi}
  \partial_{x} \varphi_{\alpha}(x) = 2\tilde \omega(x) -
  2|\Delta(x)|\sin \left(\varphi_{\alpha}(x)-\vartheta(x) \right) \;, 
\end{equation}
where we have written 
$\Delta ( x ) = | \Delta (x ) | e^{ i \vartheta ( x ) }$.
Because $|\Phi_{+} (0)| = |\Phi_{-} ( L )| =1$ for $( \omega -  V_{\infty})^2 < \Delta_{\infty}^2$,
the initial values $\varphi_{+} ( 0 )$ and $\varphi_{-} ( L )$ are
real. Hence
the solutions of Eq.(\ref{eq:phi}) remain real, which implies that 
$ | \Phi_{\alpha} ( x ) | = 1$ for all $x$. From
Eq.(\ref{initial}) we obtain for the initial values 
 \begin{equation}
 \label{phiBC}
 \tan \varphi_{+} (0) = \tan \varphi_{-} (L) =
 \frac{\omega-V_{\infty}}{\sqrt{\Delta_{\infty}^{2}-
 (\omega-V_{\infty})^{2}}}   \;.
 \end{equation}
Note that the $\varphi_{\alpha}(x)$ are
unreduced phases which are not limited to take values between $0$
and $2\pi$. 
In terms of the $\varphi_{\alpha}(x)$ the local DOS can
be written as
\begin{eqnarray}
  \label{localDOS}
  \rho(x,\omega) & = & -\frac{1}{\pi} \text{Im} \cot
  \left(\frac{\varphi_{+}(x)-\varphi_{-}(x)}{2}+i0\right) \nonumber \\
  & = & 2 \Sigma_{m = - \infty}^{\infty} \delta
  \left(\varphi_{+}(x)-\varphi_{-}(x) - 2 \pi m \right) \; .
\end{eqnarray}
We now make the $\omega$-dependence of $\varphi_{\alpha}(x)$ explicit
again. Since the right hand side of Eq.(\ref{eq:phi}) is a $2 \pi$-periodic
function of $\varphi_{\alpha}(x)$ it follows that if
$\varphi_{+}(x,\omega)-\varphi_{-}(x,\omega)=2\pi m$ for one $x$,
this must also be true for all $x$. This implies that only 
for discrete values of $\omega$ does Eq.(\ref{localDOS}) yield
a contribution to the local DOS.
We get a delta-peak contribution to the total DOS if
$\varphi_{+}(L,\omega) = 2\pi m +\varphi_{+}(0,\omega)$, where
$\varphi_{+}(0,\omega)=\varphi_{-}(L,\omega)$ is 
given by Eq.(\ref{phiBC}). Since
$\partial_{\omega} \varphi_{+}(x,\omega) > 0$, the integrated total
DOS is given by
\begin{equation}
  \label{IDOSwithBC}
  {\cal N(\omega)} = \frac{1}{L}\left[\frac{\varphi_{+}(L,\omega) -
      \varphi_{+}(L,0)}{2\pi} - C(\omega) \right]_{\text{int}} \;,
\end{equation}
where $[z]_{\text{int}}$ gives the integer value of $z$, 
and 
$C(\omega)=(\varphi_{+}(0,\omega) -
\varphi_{+}(0,0))/2\pi$ is a finite size correction
of order unity that depends on the initial condition.
For  {\em real}
$\Delta(x)$, a similar equation has been 
derived  by Lifshits, Gredeskul, and Pastur\cite{Lifshits88} within  the
phase formalism. While these authors use a rather unphysical
boundary condition, 
we can cope with
arbitrary $\Delta_{\infty}$ and $V_{\infty}$.
In the
thermodynamic limit the integrated DOS
is independently of the boundary conditions given by
\begin{equation}
  \label{IDOS}
    {\cal N(\omega)} = \lim_{L \to \infty} (\varphi_{+}(L,\omega) -
      \varphi_{+}(L,0))/2\pi L \;.
\end{equation}
For large frequencies we recover the classical high-frequency limit
${\cal N}_{0} (\omega)= \omega/ \pi$, so that the DOS  $ \rho ( \omega ) =
\partial_{\omega} {\cal N} (\omega)$ is given by $\rho_{0} =
1/\pi$. The white noise limit is also easily recovered:
in this case Eq.(\ref{eq:phi}) implies that the probability
distribution of $\varphi_{+} ( x )$ satisfies a Fokker-Planck
equation, which  was first solved by 
Ovchinnikov and Erikhman\cite{Ovchinnikov77} for the commensurate case.
For the most general case with complex $\Delta (x)$ see Ref.\cite{Hayn96}.

{\it{Numerical algorithm.}}--
In the following we present an exact algorithm which allows to
numerically calculate the
(integrated)
DOS for stepwise constant potentials.
By choosing the step size sufficiently small, arbitrarily given potentials
may be approximated in this way.
Assuming that in the open intervals $] x_n , x_{n+1} [$
the potentials $\Delta ( x )$ and $V ( x )$  are given by the
constants $\Delta_n$ and $V_n$,
the matrix $U_{+}(x)$ can be written as a
finite product of matrices of the form
\begin{eqnarray}
  \label{T}
  e^{-iM_{n} \delta_{n}}  & \equiv &
  \left(
    \begin{array}{cc}
      t_{n} & i r_{n}^{\ast} \nonumber \\
      -i r_{n} &  t_{n}^{\ast}
    \end{array}
  \right)
  = \cosh[\sqrt{|\Delta_{n}|^2-\tilde \omega_{n}^2}\delta_{n}] \sigma_{0}
  \nonumber \\
  &  &  \hspace{-17mm} +
  i \sinh[\sqrt{|\Delta_{n}|^2-\tilde \omega_{n}^2}\delta_{n}]
  \; \frac{\Delta_{n} \sigma_{+} -\Delta_{n}^{\ast} \sigma_{-} + \tilde
    \omega_{n} \sigma_{3}}{\sqrt{|\Delta_{n}|^2-\tilde \omega_{n}^2}}
    \; ,
\end{eqnarray}
where $\tilde \omega_{n} = \omega-V_{n}$ and $\delta_{n} =
x_{n+1}-x_{n}$. For the Riccati variable
$\Phi_{+} ( x )$ satisfying
Eq.(\ref{Riccati}) this implies the recurrence relation
\begin{equation}
  \Phi_{+}(x_{n+1}) = \frac{r_{n}^{\ast}+t_{n}
    \Phi_{+}(x_{n})}{t_{n}^{\ast}+r_{n} \Phi_{+}(x_{n})} \; .
    \label{eq:recurrence}
\end{equation}
We found that for a given realization of $\Delta_n$ and $V_n$
it is easier to calculate the dynamics of $\Phi_{+}(x)$ and to keep
track of its phase than to directly solve Eq.(\ref{eq:phi}). Whenever
${\rm Re} \Phi_{+}(x) > 0$ and there is a sign change in
${\rm Im} \Phi_{+}(x)$,
the winding number
$[\varphi_{+}(x) /2\pi]_{\text{int}}$ is changed by one.
To detect all such  
changes we demand that the length $\delta_{n}$ of all intervals
satisfies $2(|\tilde \omega_{n}|+|\Delta_{n}|) \delta_{n} <
\pi/2$. 
Since very long chains show a self-averaging effect,
we only need to simulate one typical  chain
to obtain the average DOS.

To generate Gaussian disorder with 
correlation length $\xi$ we have
found the following realization of an Ornstein-Uhlenbeck 
process\cite{Bartosch99c}, which is much simpler than the
algorithm proposed in Ref.\cite{Tchernyshyov98}.
Using the Box-Muller algorithm\cite{numericalrecipes},
we generate independent Gaussian random
numbers $g_n$ with 
$\langle g_n \rangle = 0$ and $\langle g^2_n \rangle = 1$.
For real $\Delta ( x )$  we set $\Delta_n = 
\Delta_{\rm av} + \tilde{\Delta}_n$ and 
generate the $\tilde{\Delta}_n$ recursively according to
\begin{equation}
\tilde{\Delta}_0  =  \Delta_{s} g_0 \; \; , \;  \;
\tilde{\Delta}_{n+1} = a_n 
\tilde{\Delta}_n
 +  \sqrt{ 1 - a_n^2}  \Delta_{s} g_{n+1}
 \; ,
 \label{eq:genial}
\end{equation}
where $a_n = e^{- | \delta_n | / \xi}$.
It is straightforward to show that this Markov process 
indeed leads to a Gaussian random process with the desired
properties.
Obvious advantages of our algorithm are that
the random variables $\Delta_n$ can be generated simultaneously
with the iteration of the  recurrence relation (\ref{eq:recurrence}),
and that $\Delta_{n+1}$ depends only on the previous
$\Delta_n$, so that the implementation of this algorithm  
requires practically no memory space.
Of course, our algorithm can also be used to
generate $V_n$,  and
in the complex case
$\text{Re}  \Delta_n$ and
$\text{Im} \Delta_n$ can be generated by replacing
$\Delta_{s}$ by $\Delta_{s}/\sqrt{2}$.

\begin{figure}
\begin{center}
\epsfxsize8.5cm 
\epsfbox{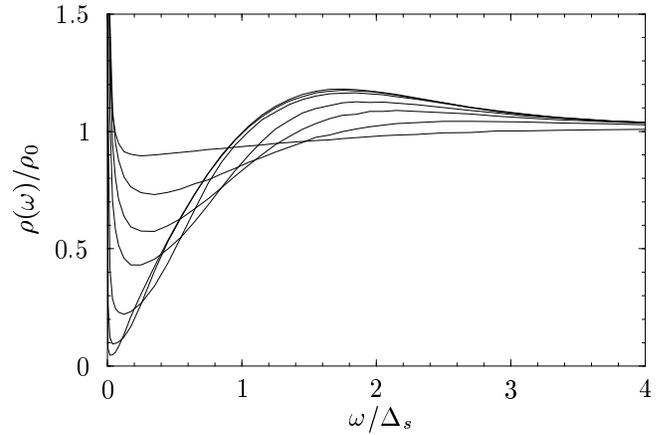}
\vspace{4mm}
\caption{Average DOS for real $\Delta (x)$ with $\Delta_s L=10^{7}$, 
  $V (x ) = \Delta_{\rm av} = 0$, and $\Delta_s \xi =
  1000$, $100$, $10$, $2$, $1$, $0.5$, $0.2$. The minimal DOS
  $\rho(\omega^{\ast})$ decreases with increasing $\Delta_s \xi$.} 
\label{fig:real}
\end{center}
\end{figure}

\begin{figure}
\begin{center}
\epsfxsize8.5cm 
\epsfbox{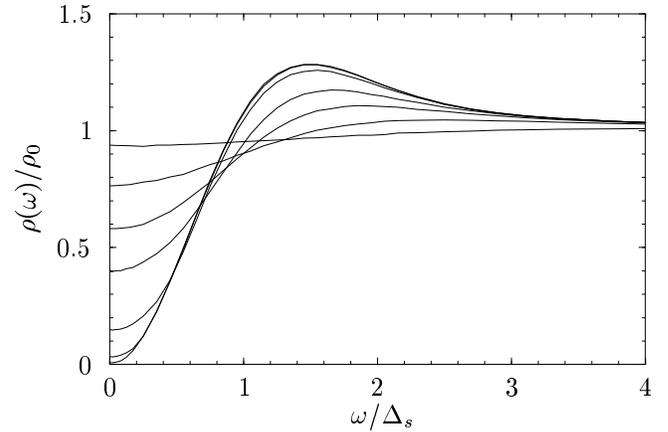}
\vspace{4mm}
\caption{Average DOS for complex $\Delta ( x)$ 
with $\Delta_s L=10^{7}$, 
$V (x ) = \Delta_{\rm av} = 0$, and $\Delta_s \xi =
  1000$, $100$, $10$, $2$, $1$, $0.5$, $0.2$. 
  $\rho(0)$ decreases with increasing $\Delta_s \xi$.}
\label{fig:complex}
\end{center}
\end{figure}

\begin{figure}
\begin{center}
\epsfxsize8.5cm 
\epsfbox{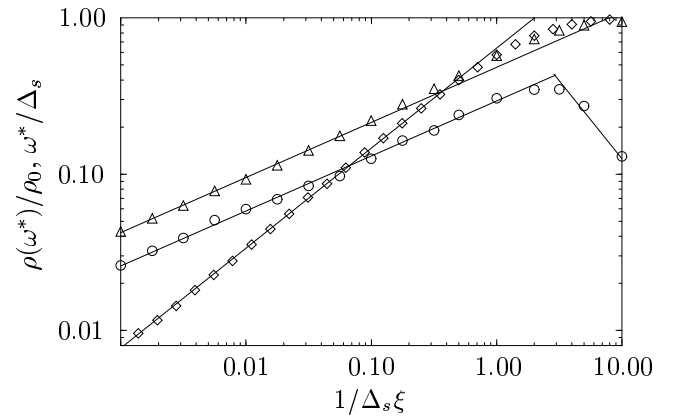}
\vspace{4mm}
\caption{Double-logarithmic plot of $\rho(\omega^{\ast})/\rho_{0}$ as
  a function of 
  $1/\Delta_{s}\xi$ for real $\Delta (x)$ (triangles) and complex
  $\Delta (x)$ (diamonds), where $\omega^{\ast}$ is the energy
  for which the DOS assumes its minimum. While $\omega^{\ast} = 0$ for
  complex $\Delta (x)$, the circles give the double-logarithmic plot
  of $\omega^{\ast}/ \Delta_s$ for real $\Delta (x)$ as a function of
  $1/\Delta_{s}\xi$.}
\label{fig:dosxidependence}
\end{center}
\end{figure}

{\it{Results.}}--
In Fig.\ref{fig:real} 
we show our numerical results for $\rho ( \omega ) / \rho_0$
for $V(x) = \Delta_{\rm av}=0$ and real $\Delta ( x )$. 
Except for $\Delta_s \xi = 1000,0.2$ we have chosen
the same values of the dimensionless parameter
$\Delta_s \xi$ as in Fig.7  
of Ref.\cite{Sadovskii79}.
One clearly sees the Dyson singularity, which exists  
for any finite value of $\xi$ and overshadows the pseudogap
at sufficiently small energies. This Dyson singularity is missed by
Sadovskii's algorithm\cite{Sadovskii79}. On the other hand,
for complex $\Delta ( x )$ this algorithm turns out to
be qualitatively correct which can be seen by comparing our data in
Fig.\ref{fig:complex} with 
those in Fig.5 of Ref.\cite{Sadovskii79}.
For a more quantitative comparison, the triangles (real $\Delta(x)$) and
diamonds (complex $\Delta(x)$) in Fig.\ref{fig:dosxidependence} show
the DOS $\rho ( \omega^{\ast} )$ at the energy $\omega^{\ast}$
where $\rho ( \omega )$ assumes its minimum. Note that in the
incommensurate case $\omega^{\ast} = 0$.
The numerical errors (which are mainly due to the finite length of the
chain) are smaller than the size of the symbols.
The straight lines are fits
to power-laws
$\rho ( \omega^{\ast} ) / \rho_0 = A ( \Delta_s \xi )^{- \mu }$.
For real $\Delta ( x )$ we obtain $A = 0.482 \pm 0.010$, $\mu = 0.3526
\pm 0.0043$,  while
for complex $\Delta ( x )$ our result is
$A = 0.6397 \pm 0.0066$ and $\mu = 0.6397 \pm 0.0024$, i.e. within
numerical accuracy we find $A=\mu$. 
The circles in Fig.\ref{fig:dosxidependence}
show  for real $\Delta (x)$ the energy scale
$\omega^{\ast}$  where $\rho ( \omega )$ is minimal.
The long solid line is a fit
to a power-law $\omega^{\ast} / \Delta_s = B ( \Delta_s \xi )^{- \gamma}$,
with $B = 0.2931 \pm 0.0074$ and $\gamma = 0.3513 \pm 0.0051$. Here we
find within numerical accuracy $\mu=\gamma$. The proportionality of 
$\rho(\omega^{\ast})$ to the energy scale $\omega^{\ast}$, which
can be 
interpreted as the width of the Dyson singularity,
can also directly be seen in Fig.\ref{fig:real}.
Finally we note that for $\Delta_s \xi 
{ \raisebox{-0.5ex}{$\; \stackrel{<}{\sim} \;$}}  0.2$ our algorithm
produces results consistent with the white noise limit $\Delta_{s} \xi
\ll 1$. While $\rho(0) \to 1$ in the incommensurate case, in the
commensurate case we obtain from the exact solution of Ovchinnikov and
Erikhman \cite{Ovchinnikov77} $\rho(\omega^{\ast})/\rho_{0} \to
0.9636$, and $\omega^{\ast} 
\to 1.2514 \Delta_{s}^{2} \xi$, which determines the short solid line in
Fig.\ref{fig:dosxidependence} describing $\omega^{\ast}(\xi)$
in the white-noise limit.

{\it{Summary.}}-- 
We have developed a powerful numerical algorithm 
to calculate the average DOS  of the FGM with very high accuracy. The 
algorithm can be used for arbitrary forward and backward
scattering potentials, so that it is not restricted to the
case of vanishing averages and Gaussian statistics which we further
considered in this work. Our main results are:
({\em{a}}) for commensurate chains 
in the absence of forward scattering
the DOS exhibits for large $\xi$ a pseudogap and a Dyson singularity.
We have explicitly calculated the width of the Dyson singularity
as a function of $\xi$. 
The most promising experimental systems to
detect Dyson singularities are
spin chains\cite{Bunder99}. 
({\em{b}}) In the incommensurate case the algorithm proposed by 
Sadovskii\cite{Sadovskii79} is qualitatively correct. However,
his result $\langle \rho ( 0 ) \rangle \propto \xi^{-1/2}$
is incorrect. This should be kept in mind for a quantitative
comparison between experimental data\cite{Schmalian98} and
calculations based on  Sadovskii's algorithm. 

We thank K. Sch\"{o}nhammer for discussions.
This work was financially supported by the
DFG (Grants No. Ko 1442/3-1 and Ko 1442/4-1).

\vspace{-0.2cm}

\vspace{-0.6cm}

\vspace{-0.6cm}


\begin{thebibliography}{99}
\bibitem{footnote1}
We measure energies relative to the Fermi energy
and set the Fermi velocity and $\hbar$ equal to unity.
\bibitem{footnoteLetter} 
In Ref.\cite{Bartosch99a}
we have calculated the exact $\rho ( x , \omega = 0^{+} )$ for
$V (x) = \Delta_{\infty} = 0$ and real $\Delta ( x )$.
In this case $\Phi_{+} ( 0 ) = \Phi_{-} ( L ) = 0$, 
and the spectrum is continuous.
\end{thebibliography}
\end{document}